# Multifunctional Wideband Digital Metasurface for Secure Electromagnetic Manipulation in S-Band

Longpan Wang, Zhuoran Zhang, Zhenyuan Li, Xuetao Gan, Xudong Bai, *Member, IEEE*,
Wen Chen, *Senior Member, IEEE*, and Qingqing Wu, *Senior Member, IEEE*

*Abstract*—Digital metasurfaces have attracted significant attention in recent years due to their ability to manipulate electromagnetic (EM) waves for secure sensing and communication. However, most reported metasurfaces operate at relatively high frequencies, primarily due to the constraints imposed by the physical scale of the dielectric substrate, thus limiting their full-wave system applications. In this work, a wideband digital reflective metasurface is presented for capable of dynamically controlling EM waves, with multifunctional applications in the lower-frequency S-band. The metasurface is composed of electronically reconfigurable meta-atoms with wideband characteristics, and designed by using trapezoidal and M-shaped patches connected by a pin diode. Simulation results show that the proposed digital metasurface could achieve wideband 1-bit phase quantization with a stable phase difference within 180°±25° and small reflection loss below 0.6 dB from 2.72 to 3.25 GHz. To validate the proposed design, a 20×20-unit metasurface array was designed, simulated and fabricated. By dynamically adjusting the coding sequence, the metasurface could enable multi-mode orbital angular momentum (OAM) beam generation, dynamic beam scanning, and precise direction finding. These capabilities support secure sensing and secure communications through high-resolution target detection and anti-jamming beam steering, as well as physical-layer security. The proposed wideband metasurface may serve as an effective candidate for enhancing spectral efficiency and security performance in radar and wireless systems.

## I. INTRODUCTION

TRADITIONAL mechanical radar systems mostly rely on antennas with fixed radiation patterns, and due to their inherent lack of flexibility, these systems often result in suboptimal performance, especially in scenarios requiring agile target tracking or adaptive interference suppression [1], [2], [3]. While the advent of phased arrays has solved such problems by being able to independently control the element excitation phase and generate focused beams for various operational conditions [4], [5], which suffer from substantial energy losses due to the feeding network mechanisms employed [6], [7]. Moreover, the complexity involved in the design, manufacturing, and maintenance of these systems further drives up costs and consumption, thus limiting their applicability in more dynamic and cost-sensitive applications [8]. These limitations become particularly critical in emerging secure sensing and communication scenarios, where high-resolution and agile beamforming are essential to enable precise target detection, conduct robust anti-jamming capabilities, and enhance the physical-layer security.

In light of the above challenges, there has been a growing interest in the development of alternative approaches that can overcome these limitations, which could offer enhanced flexibility and also reduce operational costs. The emergence of digital or programmable metasurfaces, which facilitate real-time configuration of engineered surfaces, presents a viable solution for implementing diverse functions [9], [10], [11], [12]. In 2014, digital coding metasurfaces were first proposed to create a link between the physical and digital worlds [13]. Digital metasurfaces integrated with field-programmable gate array (FPGA) could enable dynamic manipulation of electromagnetic (EM) wavefronts and also possess the advantages of low-cost and lightweight [14], [15], [16], [17]. Thereafter, active digital metasurfaces based on a variety of control mechanisms have been proposed, such as varactor diodes [18], [19], MEMS [20], and PIN diodes [21], [22]; thereamong, PIN diodes were most widely used due to the suitability for high-power RF applications and relatively low insertion loss in designing 1-bit or 2-bit reconfigurable metasurfaces [23], [24]. As for terahertz frequencies, tunable materials [25], [26] have been prevailingly proposed for the design of reconfigurable metasurfaces. However, numerous previous studies have highlighted that reconfigurable metasurfaces with narrowband characteristics are limited by their inability to efficiently cover a wide frequency range, thus reducing the flexibility and performance in systems requiring frequency tuning or adaptation. By implementing construction designs that incorporate wideband features, it is feasible to enhance the bandwidth, and techniques such as parallel placement of asymmetric monopoles [27], a metal strip of an infinite length with cross-polarized [28], and magnetoelectric dipole elements [29] have been introduced to effectively expand the overall bandwidth for digital metasurfaces.

Manuscript received        , 2025; revised         , 2025; accepted          , 2025. Date of publication             ; date of current version            . This work is supported by National Key Research and Development Program of China (2022YFA1404800), the National Natural Science Foundation of China (62471399, 62375225), the Shanghai Kewei Foundation (24DP1500500), and the Gusu Leading Talents of Innovation and Entrepreneurship (ZXL2024332). (*Corresponding authors: Xudong Bai, Qingqing Wu*)

L. Wang, Y. Chen, Z. Li, X. Gan, and X. Bai are with the School of Microelectronics, Northwestern Polytechnical University, Xi'an 710129, China (email: baixudong@nwpu.edu.cn).

Q. Wu, and W. Chen are with the Department of Electronics Engineering, Shanghai Jiao Tong University, Shanghai 200240, China (e-mail: qingqingwu@sjtu.edu.cn).

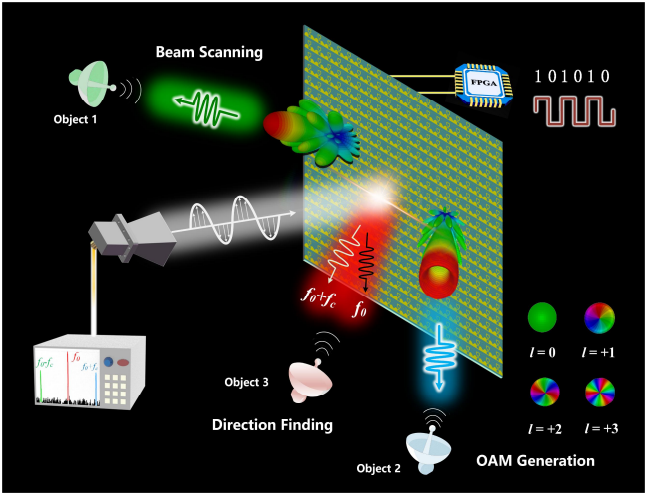

Fig. 1. Schematic diagram of the multifunctional wideband digital metasurface for high-security OAM communication, precise direction finding, as well as dynamic beam scanning.

In previous studies, most of the realized digital metasurfaces focused mainly on the reconfiguration of relatively straightforward functions [30]. For instance, a reflective digital metasurface operating in Ku-band was proposed to improve the beam scanning performance [31]. Thereafter, a series of reflective digital metasurfaces have been reported in different bands, demonstrating the relatively single function of polarization control and beam scanning, respectively [32], [33], [34]. Recently, a beam steering of reflective metasurface using a center shorted patch with two plated through holes in the Ka-band for the satellite communication [35]. Then, a 1-bit dual-polarized wheel-rudder-shaped reflective metasurface with wide-angle-scanning capability in the C-band was proposed [36]. In addition, various types of digital metasurfaces with recombined functions have also been reported, including six-channel for orthogonal circular and linear polarizations at the X-band [37], a full-space digital metasurface realizes co-polarized reflection for right-handed circularly polarized waves and cross-polarized transmission for left-handed circularly polarized waves at the X-band [38]. However, most of the above-mentioned tunable metasurfaces primarily focused on the high-frequency band, while in contrast, metasurfaces operating in the low-frequency band typically exhibit narrow bandwidth due to the constraints imposed by the physical scale of the dielectric substrate, which also leads to higher costs [39], [40].

In this work, we propose a broadband digital reflective metasurface operating in lower-frequency S-band, which is capable of simultaneously enabling high-purity multi-mode OAM generating, wide-angle beam scanning, and precise direction finding. The metasurface is composed of electronically tunable meta-atom with an integrated PIN diode, which could acquire stable phase states ($0 / \pi$) with difference of $180°\pm25°$ and very small reflection loss below 0.6 dB from 2.72 to 3.25 GHz. To validate the proposed concept, the proposed reflective digital metasurface with 20×20 elements is designed and fabricated. By modulating the real-time coding distribution of the metasurface via a programmable bias circuit, three functions can then be created. The generation of four high-purity OAM beams, dynamic beam scanning within ±60°, and precise direction finding are verified by both simulation and experiments, which demonstrates the effectiveness of the multifunctional metasurface design. The proposed integrated wideband digital metasurface may become an eligible candidate for enhancing the multiplexing capabilities and spectral efficiency of both radar and communication systems.

## II. DIGITAL METASURFACE DESIGN

The overall schematic diagram of the proposed reflective digital metasurface system for stimulating multi-mode OAM beams, wide-angle beam scanning, and precise direction finding is shown in Fig. 1. A standard horn antenna is used as the feed source, aligned along the central axis of the reflective digital metasurface to radiate the incident EM wave onto the surface. Each reflective digital meta-atom functions as a 1-bit phase shifter, modulating the incoming EM wave. For Object 1, multi-mode high-purity OAM beams are stimulated flexibly to conduct multichannel multiplexing transmission and thus improve spectrum utilization. For Object 2, dynamic beam scanning with a wide-angle coverage of ±60° is implemented to accommodate the changing environmental conditions and ensure seamless adaptation to mobile users. As for Object 3, precise direction finding is achieved by leveraging time modulation technology, which simultaneously generates fundamental and harmonic components in the received signals to facilitate accurate angle estimation. In brief, all the above-mentioned EM functions can be achieved simultaneously by the single metasurface, which not only helps to enhance the integrated multifunctional performance but also reduces maintenance costs.

Fig. 2(a)-(b) illustrate the topology of the proposed reflective meta-atom. The reconfigurable meta-atom is designed by using a resonant tunable approach for 1-bit reflection phase modulation. The structure consists of two metallic layers, a scattering layer and a metal ground plane, separated by dielectric substrates of FR-4 (blue layers) and Rogers 4003C (cyan layer). The lower FR-4 substrate has a relative permittivity of 4.3, a loss tangent of 0.004, and a thickness of 6.6 mm, while the upper Rogers 4003C substrate has a relative permittivity of 3.55, a loss tangent of 0.0027, and a thickness of 1.524 mm. The scattering layer comprises a PIN diode connected to the trapezoidal and M-shaped patches, which are etched to enhance resonance and broaden the bandwidth. An additional inductance is integrated on the side of the trapezoidal patch, and is connected to the bias line to choke the high-frequency signals, thus optimizing the overall scattering efficiency. To further minimize the impact on high-frequency performance, the biasing line is positioned at the center of the non-radiating patch edge. The optimized parameters are $p$=50, $l_1$=25.15, $l_x$=17.14, $l_y$=11.21, $W_a$=2.7, $W_1$=0.3, $W_b$=17.21, $m_a$=9, $W_2$=0.75, all in the unit of millimeters.

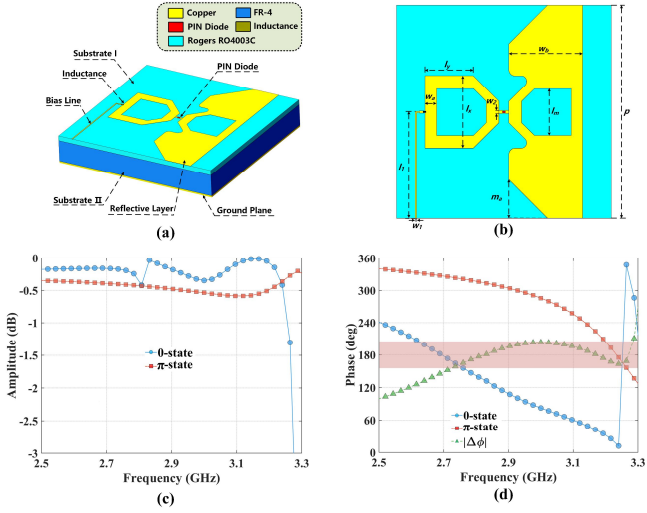

Fig. 2. (a-b) Topology of the scattering layer patch. Simulated reflection coefficient and phase of the 1-bit reflective meta-atom: (c) amplitude, and (d) phase. (The shaded area represents a phase difference of 180°±25° within the frequency range of 2.72 ~ 3.25 GHz.)

The PIN diode, MACOM MADP-000907-14020, is adopted to acquire lower insertion losses within the designed frequency band, which is modeled by the equivalent lumped components in series or parallel for the two states, respectively. For a positive DC-biasing voltage, the proposed reconfigurable meta-atom is operating at the π-state along with a typical series resistor $R_\pi$ = 5.2 Ω and inductor $L_\pi$= 100 nH adopted for the PIN diode, while for a negative DC-biasing voltage, the proposed reconfigurable meta-atom is operating at the 0-state along with a typical parallel capacitance $C_0$= 0.025 pF and inductor $L_0$ = 100 nH employed for the PIN diode. The numerical simulation of the proposed reflective meta-atom is carried out with the help of the commercial software package CST Microwave Studio by using unit cell boundary conditions along with the Floquet-port excitations. The simulated reflection coefficients of the meta-atom in both amplitude and phase for the binary states are given in Fig. 2(c)-(d). For a normal incidence EM wave, the meta-atom reflection coefficient is less than 0.6 dB from 2.72 to 3.25 GHz when the reconfigurable meta-atom is operating at both π-state and 0-state. From the 0-state, it can be observed that the reflection coefficient remains relatively stable in the entire working range band, with the reflection loss lower than 0.6 dB. In contrast, for the π-state, the reflection coefficient exhibits fluctuations, with a slight increase of loss near the central frequency of 3.0 GHz. Overall, the reflection losses for both states remain very low throughout the frequency range. Figure 1d shows the phase of the meta-atom, where denotes the absolute value of the phase difference between the two different states. The shaded gray region indicates that the phase difference of the reflective digital metasurface meta-atom lies within the range of 180° ± 25° within frequency range 2.72 to 3.25 GHz, covering a fractional bandwidth up to 17.8%. In general, the proposed reconfigurable meta-atom can achieve good 1-bit phase tuning with very low reflection losses, making it appropriate for the overall digital metasurface array design.

As a proof of concept, a wideband digital metasurface is deployed by using 20×20 meta-atoms, with an overall size of 1000 mm×1000 mm. Since each meta-atom of the metasurface is integrated with a PIN diode and an inductance, a total of 400 PIN diodes and 400 inductances are employed to provide the individual phase modulation. The top scattering layer prototype of the proposed reflective metasurface is also provided in Fig. 10 (Appendix A). To simplify the top layer design layout and minimize the bias line interference, the bias line is divided into four symmetric parts. Each section is equipped with 50 bias lines, all connected to external load devices via connectors, thus enabling real-time control of the metasurface coding. Besides, the distance from the horn to the metasurface, also known as focal length $F$, is set as 364 mm, to ensure good performance.

## III. HYBRID MULTIFUNCTIONAL REALIZATION

In practical applications, especially for dynamic communication systems and complex environments, the coordination of different functions is crucial for achieving efficient and flexible system performance. A system based on our proposed digital metasurface could capitalize on its unique advantages across various scenarios. First, in the scenario of multi-channel multiplexing transmission, the flexible excitation of multi-mode high-purity OAM beams can effectively enhance spectrum utilization, thereby improving the overall communication capacity of the system. Second, in the case of reflected wave reception and dynamic beam scanning, the proposed digital metasurface could adapt to environmental changes by providing wide-angle beam scanning and directional functionality, thus ensuring stable signal transmission in mobile user environments. Finally, in applications involving precise positioning and target tracking, the system can achieve accurate localization and target recognition through precise direction finding and phase control. Through the coordination of these three functions, the proposed digital metasurface could not only improve the spectral efficiency, but also offer flexible communication and positioning services during complex and changing environments, thus meeting the demands of modern communication, intelligent transportation, and Internet of Things (IoT) applications.

### A. OAM Generation

OAM beams are considered a promising technology for modern wireless communications, offering not only spatial multiplexing capabilities for high-capacity transmission but also enhanced security. The unique mode structure of OAM waves requires precise alignment between the transmitted and received modes, which inherently enhances the level of encryption as well as security. Efforts have been dedicated to leveraging this technology to maximize channel capacity. Nowadays, digital metasurfaces provide innovative solutions for generating multi-mode OAM beams due to their agile EM modulation capability [41], [42], [43]. The dynamic OAM beams have a rotational phase profile associated with an

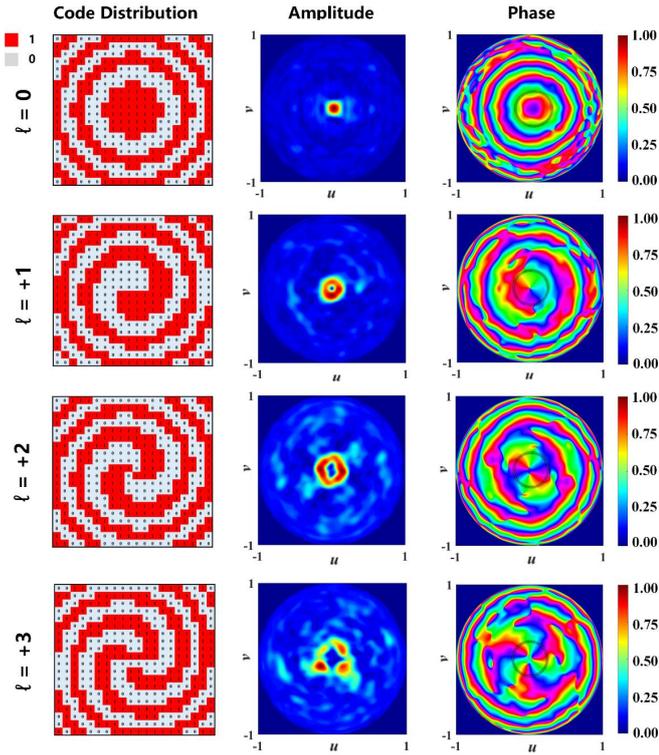

Fig. 3. Code distributions and simulated radiation patterns for four OAM modes $\ell = 0$, $\ell = +1$, $\ell = +2$ and $\ell = +3$.

azimuth angle, which is

$$\varphi(x, y) = \ell \cdot \arctan(y/x) \quad (1)$$

where $(x, y)$ is the unit position coordinates on the reflective metasurface meta-atom. To stimulate the converged multi-mode OAM beams through the programable reflective metasurface, the optimal compensation phase for each unit should be determined based on the following formula

$$\Phi_{OAM}(x, y) = 2\pi\left(\sqrt{(x^2+y^2)+F^2} - F\right)/\lambda + \varphi(x, y) \quad (2)$$

where $\ell$ is the designed mode number, $\lambda$ is the free-space wavelength at the designed frequency, and $F$ is the focal length of the feed phase center. Moreover, since only binary phase states are available for the reconfigurable meta-atom, the compensation phases should be furtherly quantized into two coding states according to

$$\Phi_q = \begin{cases} 0, & \phi_{OAM} \in [0+2n\pi, \pi+2n\pi) \\ \pi, & \phi_{OAM} \in [\pi+2n\pi, 2\pi+2n\pi). \end{cases} \quad (3)$$

To perform a preliminary verification of the proposed reflective digital metasurface, the generation of four distinct OAM beams with topological charges $\ell = 0$, $\ell = +1$, $\ell = +2$, and $\ell = +3$ is numerically investigated, as shown in Fig. 3. The optimal quantized code distributions for these four modes, calculated according to Equation (3), are presented in the first column of Fig. 3. From these distributions, the corresponding helical patterns are clearly observed, with the mode number corresponding to the number of helices. The second column displays the simulated far-field patterns for the four OAM beams, where high-intensity vortex patterns are observed, and a high-gain directional pencil beam is generated for the mode $\ell=0$. The far-field phase distributions for all four OAM modes are shown in the third column of Fig. 3. The simulation results clearly reveal that the vortex patterns for these four OAM modes distinctly showcase the characteristic helical wavefronts of OAM beams.

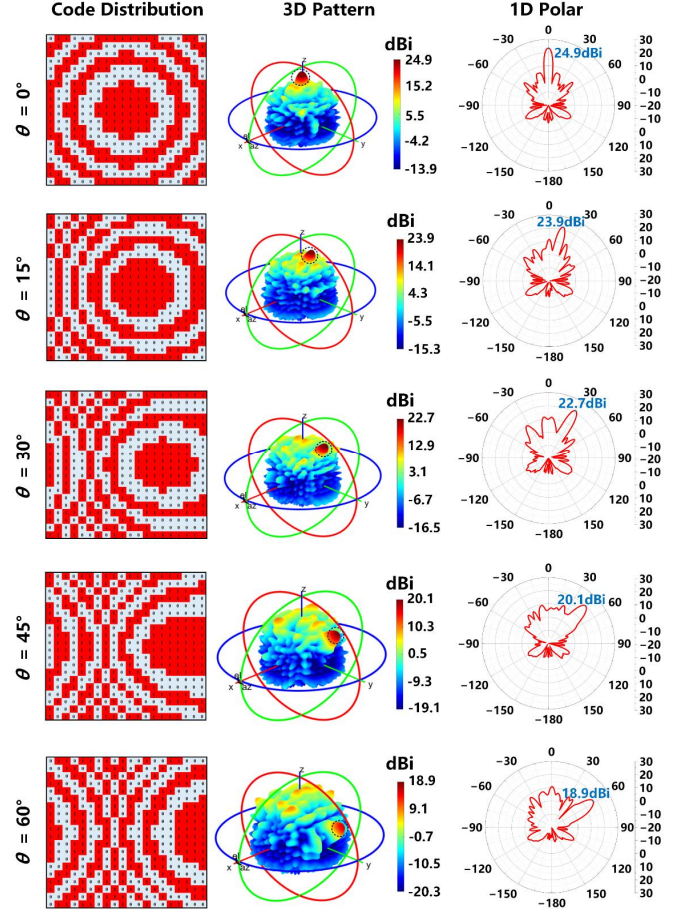

Fig. 4. Code distribution and simulated patterns for five scanning angles $\theta = 0°$, $\theta = 15°$, $\theta = 45°$ and $\theta = 60°$.

*B. Dynamic Beam scanning*

Beam scanning is an indispensable technique for modern radar and communication applications. The ability to perform beam scanning electronically could enhance the flexibility and scalability of the overall systems, thus making the digital metasurface an attractive candidate for advanced beam management in dynamic environments.

As for a feed source that is centrally located at a certain distance from the surface of the reflective, the EM field incident at an angle on each array element may be locally considered as a plane wave. To generate a scanning pencil beam in the desired direction, the phase compensation for the reflective array meta-atoms should satisfy the following formula

$$k_0\left(R_i - \overline{r_i} \cdot \hat{r_0}\right) - \Delta\phi_{mn} = 2n\pi \quad n \in Z \quad (4)$$

where $k_0$ is the free-space wavenumber, $R_i$ is the distance from the feed horn to the $i^{th}$ element, $\overline{r_i}$ is the position vector of the $i^{th}$ element, and $\hat{r_0}$ is the desired direction of the pencil beam.

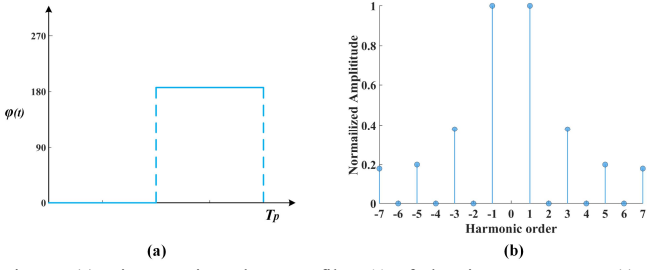

Fig. 5. (a) Time-varying phase profile $\varphi(t)$ of the time sequence $g_n(t)$ to achieve modulation. (b) Normalized coefficient amplitude of the generated harmonics.

To evaluate the beam scanning performance of the proposed reflective digital metasurface, Fig. 4 presents the simulated scattering patterns for a scanning range from $\theta = 0°$ to $60°$ with a step of $15°$. The optimal quantized code distributions for the five directional beams are generated using Equation (4), which can be simply controlled by the modulation of the ON/OFF states of the PIN diode. In the 3D pattern, a distinct scanning pencil beam is observed, progressively shifting from $0°$ to $45°$ while maintaining high beam intensity and low sidelobe levels. As the scanning angle increases, the phase difference between adjacent meta-atoms grows, causing the energy to spread in other directions and resulting in the formation of sidelobes. The polar patterns further illustrate the angular resolution and beamwidth characteristics as the beam scans across the specified range. It is worth noting that the proposed digital metasurface achieves stable beam focusing within the entire scanning range, thereby validating its suitability for wide-angle beam control applications. Simulation results demonstrate that the proposed digital metasurface excels in both beam quality and scanning range, offering significant advantages for dynamic beamforming in radar, communication, and sensing systems.

*C. Direction finding*

Direction finding is also a critical functionality in radar tracking and radio navigation systems. Especially, real-time direction finding enabled by reconfigurable metasurfaces can be utilized for anti-jamming communications, allowing the detection of intrusion sources and interference directions at the physical layer for proactive mitigation to ensure secure communications [44], [45]. Conventional antenna direction finding typically relies on multiple arrays or the rotation of a single antenna to determine the direction of the signal source. These methods estimate the incident angle of the signal by utilizing information such as signal strength, phase difference, or time delay, along with the positional relationships of the antenna arrays [46], [47]. However, these traditional approaches are often limited by antenna configuration, sensitivity, and environmental factors, and require complex hardware. In contrast, the proposed direction-finding method based on digital metasurfaces incorporates an initial phase compensation technique. In particular, during the direction-finding process, the phase compensation effectively corrects phase discrepancies caused by varying incident angles, thereby improving the accuracy and reliability of the direction detection. Suppose a metasurface consists of M×N meta-atoms, a far-field radio frequency (RF) signal with a carrier frequency fc irradiates the metasurface by direction $(\theta,\varphi)$. The direction vector of the incident signal can then be written as

$$r = (\sin\theta\cos\varphi, \sin\theta\sin\varphi, \cos\theta) \quad (5)$$

where $\theta$ is the angle between the line connecting the object to the origin of metasurface and the $+z$ axis, and $\varphi$ denotes the angle between the line connecting the projected point of the object on $xOy$ plane to the origin and the $+x$ axis. Owing to the low back-lobe level of the horn antenna, it can be assumed that all the received signals are solely from the reflection of the metasurface. The horn antenna is positioned at $[0,0,F]$, where $F$ represents the focal point. The distance between adjacent meta-atoms of the metasurface along x and y-axes is denoted by $P$. Consequently, the distance between the $(m,n)$-th meta-atom and the horn antenna can be expressed as

$$L_{m,n} = \sqrt{(m-\frac{M+1}{2})^2 P^2 + (n-\frac{N+1}{2})^2 P^2 + F^2} \quad (6)$$

To estimate the 2D direction, the metasurface is divided into subarrays. The modulation sequence $U_{m,n}(t)$ applied to $(m, n)$th meta-atom is given

$$U_{m,n}(t) = \sum_{q=-\infty}^{\infty} g_n(t - qT_p) \quad (7)$$

where

$$g_n(t) = \begin{cases} 1, \frac{n-1}{M}iT_p < t \leq \frac{n}{M}iT_p \\ -1, others \end{cases} \quad (8)$$

and $T_p$ is the time period; besides, "1" means the phase shift is $180°$, while "-1" means phase shift of $0°$, and $i$ is the index of the modulation period.

Considering the properties of periodic functions, the time sequence is decomposed by using the Fourier series

$$U_{m,n}(t) = \sum_{h=-\infty}^{+\infty} a_h e^{j2\pi F_p t}$$
$$a_h = \frac{1}{T_p}\int_0^{T_p} U_{m,n}(t)e^{-j2\pi kF_p t}dt \quad (9)$$

where $F_p$ is the modulation frequency and equal to $1/T_p$, and the Fourier coefficient $a_h$ represents the $k^{th}$ harmonic generated in the $n^{th}$ element.

To streamline the implementation, we have excluded the possibility of absorption states. Consequently, the reflection coefficient becomes time-varying, taking only two values "+1" and "−1", which correspond to the reflection phases of "0" and "π", respectively, as shown in Fig. 5(a) within a period of $T_p$. Fig. 5(b) illustrates the relationship between the amplitude of the harmonic coefficients and the harmonic order.

From an engineering perspective, the patterns generated by the subarray are similar, with their phase centers spaced by half a wavelength. Therefore, during the lateral and vertical modulation phase, the ratio of the first harmonic component to the fundamental component can be derived from the following formula

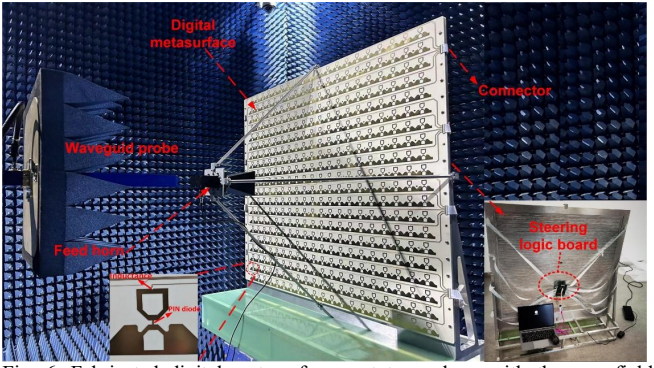

Fig. 6. Fabricated digital metasurface prototype along with the near-field measurement environment.

$$\beta_x = \frac{AF(\theta,\varphi,1)}{AF(\theta,\varphi,0)} = -\frac{2(j+1)}{\pi}\tan\left(\frac{\pi D}{\lambda}\sin\theta\cos\varphi\right)$$
$$\beta_y = \frac{AF(\theta,\varphi,1)}{AF(\theta,\varphi,0)} = -\frac{2(j+1)}{\pi}\tan\left(\frac{\pi D}{\lambda}\sin\theta\sin\varphi\right)$$
(10)

According to the Equation (10), the estimated result of $\theta$ and $\varphi$ can then be obtained, which can be written as

$$\theta_{est} = \arcsin\frac{\lambda\arctan\left(-\pi R_2/(2+2j)\right)}{2\pi D\sin\varphi_{est}}$$
(11)

and

$$\varphi_{est} = \begin{cases} \arctan\dfrac{\arctan\left(-\pi R_2/(2+2j)\right)}{\arctan\left(-\pi R_1/(2+2j)\right)}, & \sin\theta\cos\varphi > 0 \\ \pi + \arctan\dfrac{\arctan\left(-\pi R_2/(2+2j)\right)}{\arctan\left(-\pi R_1/(2+2j)\right)}, & \sin\theta\cos\varphi < 0 \end{cases}$$
(12)

## IV. Experimental Verification

To further inspect and validate the functionality of our proposed digital metasurface for dynamically generating multi-mode OAM beams, wide-angle beam scanning, and precise direction finding, a prototype metasurface array with 20×20 meta-atoms was fabricated. As depicted in Fig. 6, the metasurface was first measured and characterized using a three-dimensional platform within a near-field anechoic chamber for OAM beams and beam scanning. During the measurements, an open-ended waveguide probe, positioned two wavelengths behind the horn feed, was employed as the receiving antenna. The probe is connected to a port on the network analyzer to acquire near-field data. The reflective digital metasurface is illuminated by a linear polarization standard waveguide horn (10-dBi gain), which is positioned in the central axis of the metasurface and connected with the other port of the network analyzer.

To minimize interference during the measurement process, the control logic board was positioned at the rear of the metasurface, as shown in the lower right corner. The FPGA, is adopted as the main processing system to regulate the control signal in parallel for all 400 PIN diodes on the array surface within the assigned clock signals, according to the optimal quantized code distributions. For the π-state, the steering logic board provides a 10mA bias current and a +1.3 V positive DC bias voltage to the PIN diodes. In contrast, for the 0-state, a negative DC-biasing voltage was applied with negligible current. The measurement values of OAM beams and beam scanning are accurately calculated by using the Fourier transform integration method from planar near-field measurement data, which converts near-field data into far-field information to assess the beam distribution and propagation characteristics [48].

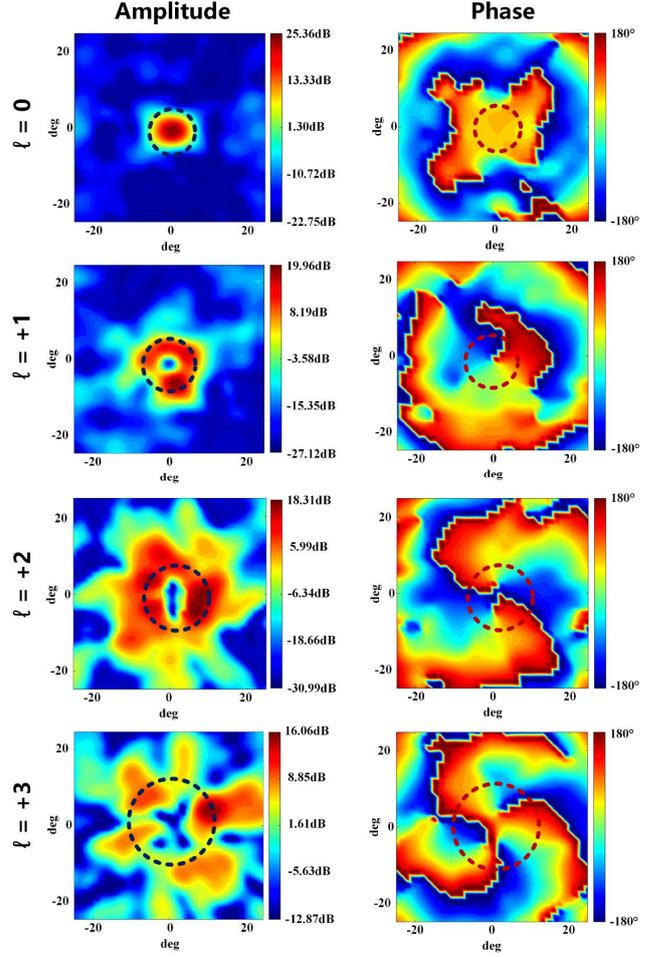

Fig. 7. Experimental patterns of the fabricated metasurface for four OAM modes, including both amplitude and phase.

Fig. 7 shows the measured results of amplitude and phase distributions for the four OAM modes of $\ell = 0$, $\ell = +1$, $\ell = +2$, and $\ell = +3$ at 3 GHz. From the amplitude patterns, it can be observed that, as the OAM mode number increases, the spatial area of the non-radiative region also expands. For OAM mode $\ell=+1$, the amplitude pattern shows a central point surrounded by a nearly ring-shaped region. In contrast to mode $\ell=+1$, the mode $\ell=+2$ exhibits a larger coverage area, with the white dashed circle highlighting the position of the main lobe. Additionally, for OAM mode $\ell=+3$, a more complete green region is surrounded by a red region, demonstrating an improved amplitude distribution. The phase distribution diagrams reveal that the number of spiral distributions increases with the mode number. As the mode number increases, the vortex patterns corresponding to the OAM modes become more clearly discernible in the phase diagram, which is likely due to the gradual avoidance of source obstruction, thereby providing a more detailed visualization of the mode structures.

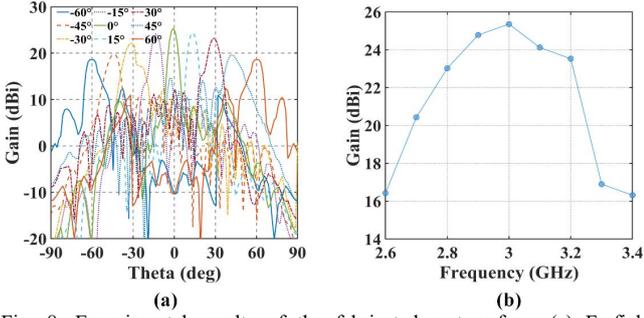

Fig. 8. Experimental results of the fabricated metasurface. (a) Farfield radiation patterns for nine different scanning angles. (b) Measured peak gains for different frequencies.

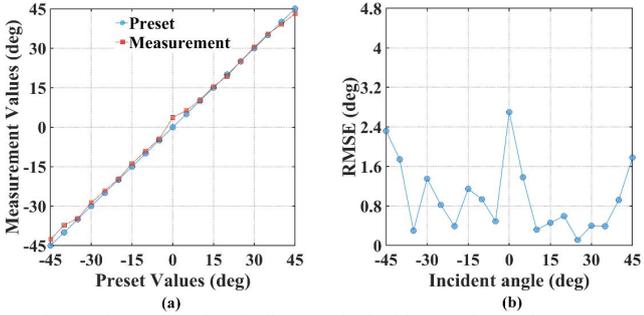

Fig. 9. Results of direction finding for the incident angle ranging from −45° to +45°. (a) The comparison of preset values and the actual measurement results. (b) RMSEs of direction finding for the test angles.

Fig. 8(a) presents the measured far-field wide-angle beam scanning for the five angles of $\theta$ = 0°, 15°, 30°, 45° and 60°. As observed, the collimated beam gain is as high as 25.36 dBi for $\theta$ = 0° at 3.0 GHz. Thereafter, the gain becomes progressively smaller as the angle continues to increase due to factors such as focal length and scanning coverage. Overall, the beam scanning exhibits well-defined radiation patterns across a wide range from $\theta$ = −60° to $\theta$ = +60°. Fig. 8(b) presents the comparative gain results at different frequencies for the scan angle of 0°, with a frequency step of 0.1 GHz. As shown in the figure, the gain increases with frequency until reaching its peak at 3.0 GHz, after which it decreases sharply with further frequency increase. This decrease in gain can be attributed to the fact that the coding distribution is designed for the central frequency 3.0 GHz. As a result, mismatching may occur between the digital reflective metasurface and the feed source at other frequencies, thus the propagation and focusing of the beam become more challenging. When the operation frequency is out of the designed range, the stimulated beams may experience greater divergence, leading to reduced directionality as well as a lower gain.

Finally, to validate the direction-finding performance of the digital metasurface, a far-field test environment was rebuilt in the microwave anechoic chamber. The specific experimental setup is shown in Fig. 11 (Appendix B). The digital metasurface was placed 20 m away in the chamber, and a 3.0 GHz single-carrier signal was transmitted from the far field. After being reflected by the digital metasurface, the signal was received by the feed antenna and processed to estimate the direction of the far-field signal source relative to the metasurface. The incidence direction of the far-field signal was preset from −45° to +45° with steps of 5°. Fig. 9(a) demonstrates the preset and measured values, and it can be seen that the measured values basically maintain a steady state with the preset values. Fig. 9(b) shows the root-mean-square errors (RMSEs) of direction finding for the test angles, which is calculated according to the following equation

$$\text{RMSE}(\theta) = \sqrt{\frac{1}{3}\sum_{i=1}^{3}\left(\theta - \hat{\theta}\right)^2}. \quad (13)$$

As shown in Fig. 9(b), the measurement absolute error is less than 2.7° from −45° to 45°. Notably, the largest test error occurs at 0°, which can be attributed to partial occlusion of the reflective digital metasurface by the feed source. The occlusion introduces perturbations in the reflected wavefront, thereby impacting the accuracy of the angle estimation.

Table I further shows the comparison of the different properties of our proposed digital metasurface and some previous works. It is evident that the designed digital reflective metasurface requires only a single active tunable component to operate at lower frequencies, offering a relative bandwidth superior to that of other existing low-frequency digital metasurfaces. More importantly, the proposed metasurface could support one-to-many functional applications, enabling high-purity OAM generation, dynamic beam scanning, and precise direction finding.

TABLE I
COMPARISON OF THE PROPOSED METASURFACE WITH SOME PREVIOUS WORKS

| Ref | Frequency (GHz) | Phase resolution | Bandwidth | Functions |
|---|---|---|---|---|
| [33] | 4.5-4.7 | 1 bit | 4.3% | Beam scanning |
| [34] | 3.05-3.25 | 2 bit | 6.25% | Intelligent imaging |
| [36] | 4.8-5.4 | 1 bit | 11.8% | Beam scanning |
| [39] | 3.5-3.8 | 1 bit | 8.2% | Signal coverage enhancement |
| [40] | 3.7-4.0 | 2 bit | 7.7% | Beam scanning Polarization conversion |
| **This work** | **2.72-3.25** | **1 bit** | **17.8%** | **OAM generation Beam scanning Direction finding** |

## V. CONCLUSION

In summary, a broadband digital reflective metasurface operating in lower-frequency S-band is proposed to realize the one-to-many secure transceiver system through integrating multiple functionalities, including multi-mode OAM generating, wide-angle beam scanning, and precise direction finding. The designed meta-atom could achieve wideband 1-bit phase quantization with a stable phase difference within 180°±25° and small reflection loss below 0.6 dB from 2.72 to 3.25 GHz, achieving a fractional bandwidth up to 17.8%. A 20×20-unit digital metasurface prototype is fabricated to evaluate its performance in terms of directional gain stability within the operating frequency band. The metasurface could enable the dynamic generation of four-mode adjustable OAM beams, including $\ell = 0$, $\ell = +1$, $\ell = +2$, and $\ell = +3$, demonstrating the versatility and effectiveness of the proposed design. Additionally, the proposed metasurface could conduct dynamic beam scanning across a wide range from $\theta = -60°$ to $\theta = +60°$. Ultimately, precise direction finding has also been successfully verified, with measured results showing good agreement with the theoretical predictions. The proposed integrated wideband digital metasurface may become an eligible candidate for enhancing the multiplexing capabilities and spectral efficiency as well as security performance of both radar and communication systems.

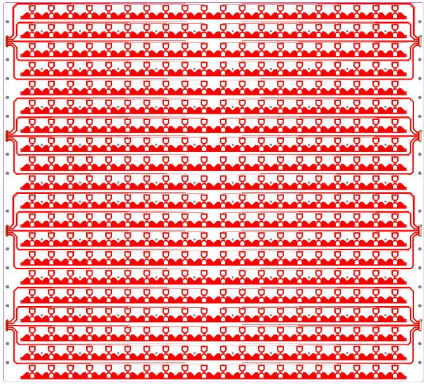

Fig. 10. Overall top layer configuration of the proposed reflective digital metasurface with 20×20 meta-atoms.

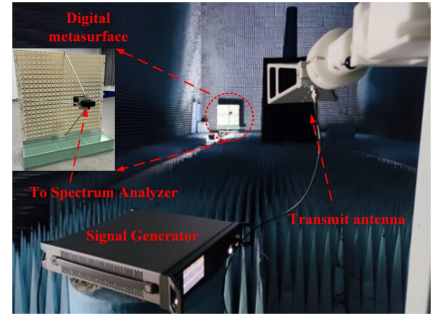

Fig. 11. Test environment for direction finding.

## APPENDIX

**Appendix A: Array design of reflective digital metasurface**

Fig. 10 shows the prototype of digital metasurface is deployed by using 20×20 meta-atoms, with an overall size of 1000 mm×1000 mm. Each meta-atom of the reflective metasurface is integrated with a PIN diode and an inductance, a total of 400 PIN diodes and 400 inductances. To simplify the top layer design layout and minimize the bias line interference, the bias line is divided into four symmetric parts. Each section is equipped with 50 bias lines, all connected to external load devices via connectors, thus enabling real-time control of the metasurface coding. The metasurface is fabricated using multilayer PCB (Printed Circuit Board) manufacturing technology.

**Appendix B: Test Environment for Direction Finding**

Direction finding performance of the reflective digital metasurface is validated in a microwave anechoic chamber, as shown in Fig. 11. The digital metasurface is placed in an appropriate location (20 meters away), with one end of its port connected to a spectrum analyzer. The standard horn antenna, serving as the transmit source, is positioned in the far-field region of the metasurface array and connected to the signal source. It is important to note that since the standard horn antenna can only receive electric fields polarized in the same direction, to accurately obtain the near-field data from the array antenna, it is essential to ensure that the polarization of the open waveguide matches that of the array. First, a single-carrier signal with a frequency of 3.0 GHz is transmitted from the transmitting antenna, and after being reflected by the metasurface, the signal was received by an antenna and processed to estimate the direction of the far-field signal source relative to the intelligent metasurface.

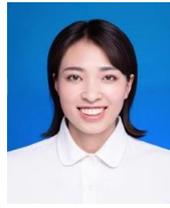

**Longpan Wang** received the B.S. degree in communication engineering from Xiangnan College (XNU), China, and the M.S. degree in electronic science and technology from Yantai University (YTU), China, in 2015 and 2019, respectively, and she is currently pursuing the Ph.D. degree in integrated circuit science and engineering from Northwestern Polytechnical University (NPU), China, in 2022 until now. Her current research interests include electromagnetic metasurfaces.

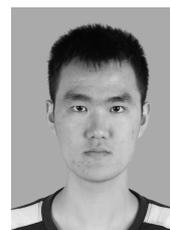

**Zhuoran Zhang** received the B.S. degree in microelectronic science and engineering from Northwestern Polytechnical University (NPU), Xi'an, China, in 2022 and is pursuing the M.S. degree in Artificial Intelligence from Northwestern Polytechnical University (NPU), Xi'an, China, until now. His current research interests include electromagnetic metasurfaces, low-profile phased arrays, reconfigurable antennas and phased array antenna.

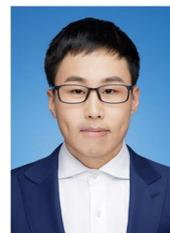

**Zhenyuan Li** received the B.S. degree in Applied Physics from Taiyuan University of Technology (TYUT), Taiyuan, China, in 2016, and pursued the M.S. degree in Electronic Science and Technology from Northwestern Polytechnical University (NPU), Xi'an, China, in 2023, until now. His current research interests include phased array antennas and space-time coding metasurfaces.

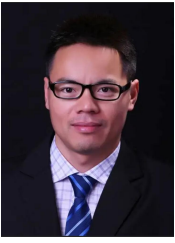

**Xuetao Gan** received the Ph.D. degree in optical engineering from Northwestern Polytechnical University (NPU), Xi'an, China, in 2013. During 2010 and 2012, he visited Columbia University, New York, NY, USA, as a joint Ph.D. student.

In 2014, he started his professional career in NPU as an Associate Professor. He is currently a Professor with the School of Microelectronics, Northwestern Polytechnical University, Xi'an, China. His research interests include light-matter interaction and device applications in nanophotonic devices. He has published more than 30 papers in journals including Nature Photonics, Nano Letters, IEEE Journal of Selected Topics in Quantum Electronics, Applied Physics Letters, Optics Express, and Optica.

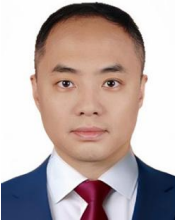

**Xudong Bai** (Member, IEEE) received the B.S., M.S., and Ph.D. degrees in electronic science and technology from Shanghai Jiao Tong University (SJTU), Shanghai, China, in 2009, 2012, and 2016, respectively.

He is currently an Associate Professor with the School of Microelectronics, Northwestern Polytechnical University (NPU), Xi'an, China. He has authored or co-authored more than 80 papers and includes one ESI highly cited paper. He also holds more than 18 patents in antennas and metamaterial technologies. His current research interests include electromagnetic digital metasurfaces, low-cost or ultrawideband phased arrays, and OAM-EM wave propagation and antennas design.

Dr. Bai received the Gusu Leading Talents Award of Innovation and Entrepreneurship in 2024. He also won the Excellent Scientific Paper Award from Suzhou Government from 2022 to 2023.

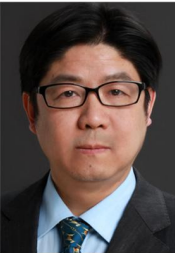

**Wen Chen** (Senior Member, IEEE) received BS and MS from Wuhan University, China in 1990 and 1993 respectively, and the Ph.D. degree from University of Electro-communications, Japan in 1999.

He is now a tenured Professor with the Department of Electronic Engineering, Shanghai Jiao Tong University, China, where he is the director of Broadband Access Network Laboratory. He is a fellow of Chinese Institute of Electronics and the distinguished lecturers of IEEE Communications Society and IEEE Vehicular Technology Society. He is the Shanghai Chapter Chair of IEEE Vehicular Technology Society, a vice president of Shanghai Institute of Electronics, Editors of *IEEE Transactions on Wireless Communications*, *IEEE Transactions on Communications*, *IEEE Access* and *IEEE Open Journal of Vehicular Technology*. His research interests include multiple access, wireless AI and reconfigurable intelligent surface enabled communications. He has published more than 180 papers in IEEE journals with citations more than 10,000 in Google scholar.

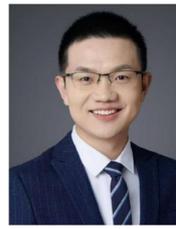

**Qingqing Wu** (Senior Member, IEEE) is currently an Associate Professor with Shanghai Jiao Tong University (SJTU). He has co-authored more than 100 IEEE journal articles with more than 40 ESI highly cited articles and more than ten ESI hot articles, which have received more than 31,000 Google citations. He has been listed as the Clarivate ESI Highly Cited Researcher since 2021, the Most Influential Scholar Award in AI-2000 by Aminer since 2021, Worlds Top 2% Scientist by Stanford University since 2020, and the Xiaomi Young Scholar. His current research interests include intelligent reflecting surface (IRS) and MIMO transceiver design.

Dr. Wu was a recipient of the IEEE ComSoc Fred Ellersick Prize, the Best Tutorial Paper Award in 2023, the Asia–Pacific Best Young Researcher Award, the Outstanding Paper Award in 2022, the Young Author Best Paper Award in 2021 and 2024, the Outstanding Ph.D. Thesis Award of China Institute of Communications in 2017, the IEEE ICCC Best Paper Award in 2021, and the IEEE WCSP Best Paper Award in 2015. He served as the Chair for the IEEE ComSoc Young Professional AP Committee and the IEEE VTS Drone Committee. He is the Workshop Co-Chair of IEEE ICC 2019–2023 and IEEE GLOBECOM 2020. He serves as the Workshops and Symposia Officer for Reconfigurable Intelligent Surfaces Emerging Technology Initiative and Research Blog Officer of Aerial Communications Emerging Technology Initiative. He was an Exemplary Editor of *IEEE Communications Letters* in 2019 and an exemplary reviewer of several IEEE journals. He serves as an Associate/Senior/Area Editor for *IEEE Transactions on Wireless Communications*, *IEEE Transactions on Communications*, *IEEE Communications Letters*, and *IEEE Wireless Communications Letters*. He is the Lead Guest Editor of *IEEE Journal on Selected Areas in Communications*.